\documentclass[pra,twocolumn,aps,superscriptaddress,showpacs]{revtex4}
\usepackage{epsfig}
\usepackage{graphicx}
\usepackage{amsmath}
\usepackage{amscd}

\newcommand{\beq}{\begin{eqnarray}}
\newcommand{\eeq}{\end{eqnarray}}
\def\be{\begin{equation}}
\def\ee{\end{equation}}
\def\ba{\begin{eqnarray}}
\def\ea{\end{eqnarray}}

\begin{document}

\title
{Converting Zitterbewegung Oscillation to Directed Motion}
\author{Qi Zhang}
\affiliation{Centre of Quantum Technologies and Department of
Physics, National University of Singapore, 117543, Singapore}

\author{Jiangbin Gong}\email{phygj@nus.edu.sg}
\affiliation{Department of Physics and Centre for Computational
Science and Engineering, National University of Singapore, 117542,
Singapore}
 \affiliation{NUS Graduate School for Integrative Sciences
and Engineering, Singapore 117597, Republic of Singapore}

\author{C.H. Oh} \email{phyohch@nus.edu.sg}
\affiliation{Centre of Quantum Technologies and Department of
Physics, National University of Singapore, 117543, Singapore}
\affiliation{Institute of Advanced Studies, Nanyang Technological
University, Singapore 639798, Republic of Singapore}

\date{\today}
\begin{abstract}
Zitterbewegung oscillation (ZB), namely, the jittering center-of-mass motion predicted by
free-space Dirac (or Dirac-like) equations, has been studied in several different contexts.
%To gain more insights into driven Dirac-like equations as well as the nature
%of ZB,
It is shown here that ZB can be converted to directed center-of-mass
motion by a modulation of the Dirac-like equation, if the modulation is on resonance with the ZB frequency.
Tailored modulation may also stop, re-launch or even reverse the directed motion of a wavepacket
with negligible distortion. The predictions may be examined by current ZB experiments using
trapped-ion systems.
\end{abstract}
\pacs{03.75.-b, 32.80.Qk, 71.70.Ej, 37.10.Vz} \maketitle

\section{introduction}

Zitterbewegung motion (ZB) originally refers to the trembling motion of a free relativistic particle (e.g., electron) as described by the Dirac equation~\cite{Schrodinger}.
%(it even does not exist when considering the quantum field theory
%\cite{Krekora}),
ZB in this context has an extremely large frequency and a very small amplitude, so it is hardly
observable. Nevertheless, the physics of ZB becomes highly relevant to a number of interesting
situations where the quantum dynamics is governed by artificial Dirac or Dirac-like equations. Examples include band
electrons in graphene \cite{Novoselov}, cavity electrodynamics~\cite{larson}, ultracold atoms in designed laser fields~\cite{JuzeliunasPRA2008,VaishnavPRL2008,MerklEPL2008,Zhangannal,ZhangPRAzb}, macroscopic
sonic crystals and photonic superlattices~\cite{ZhangPRL2008}, as
well as trapped-ion systems~\cite{LamataPRL,GerritamaNature}.  For a review of recent fruitful studies of ZB physics using electrons in
semiconductors, see Ref. \cite{zbreview}.

Many quantum dynamical phenomena induced by a control field
have been explored using the Schrodinger equation. Much less is known in the case of
driven Dirac or Dirac-like equations. This work is concerned with ZB in a simple version of a driven
Dirac-like equation. Previously, we showed that ZB in the cold-atom context as a quantum coherence phenomenon can be coupled with some mechanical oscillations, resulting in the control of the amplitude, the
frequency, and the damping of the ZB associated with a wavepacket~\cite{ZhangPRAzb}.
Going one step further, it should be of interest to study what happens to ZB if one term in a Dirac-like equation
is modulated with a frequency on-resonance with the ZB frequency.  Indeed, in both classical mechanics
and quantum mechanics, it is always expected that an oscillation phenomenon can behave remarkably different
when it is subject to an on-resonance driving field.  A latest example is Bloch oscillation subject to on-resonance driving,
where Bloch oscillation may be converted to ``super-Bloch oscillation"~\cite{Bloch1}
or even to directed motion~\cite{NewJP}.

In this work we start from a Dirac-like equation that is already experimentally realized in the context of trapped-ion systems~\cite{GerritamaNature}.
We then introduce a modulation to the Dirac-like equation, whose frequency is on resonance with the ZB frequency.
Due to the on-resonance driving, it is shown, both analytically and numerically, that ZB associated with a wavepacket can be converted to directed wavepacket
motion. By tailoring the modulation, we may also stop, re-launch or even reverse the directed motion of a wavepacket
with negligible distortion.  These results indicate that ZB is not a mysterious oscillation, and it provides an extra useful
control knob for manipulating quantum motion.

%We show both
%theoretically and numerically that, in the resonant case with
%driving frequency equaling the original ZB frequency, ZB will
%transit to a directed motion in real space. The scheme is an
%analogue of driving Bloch oscillation, where the Bloch oscillation
%will increase its amplitude to super Bloch oscillation \cite{Bloch1}
%and even transit to directed motion \cite{NewJP}. The great
%advantage of the driving ZB lies in that no periodic potential or
%lattice is needed. It is shown that the wavepacket can be
%effectively manipulated via appropriate driving, including starting
%the directed motion, suspending the directed motion, as well as
%relaunching it in either the original direction or the opposite
%direction, without intervening the wavepacket directly.

%The possible experimental implementation of directed motion via
%driving ZB is also discussed. We put the emphasis on the trapped-ion
%system since ZB in ion-trap-laser system has already been
%experimentally implemented \cite{GerritamaNature}. Our driving ZB
%scheme takes advantage of the great power of ZB-simulation system,
%where the essential parameters of Dirac-like equation can be
%experimentally adjusted to achieve more favorable ZB frequency and
%amplitude. Due to the quick developing of the ZB studies and its
%ubiquitousness, our scheme will shed more light on the ZB subject
%and offer new insight into it.

\section{theoretical analysis}

Let us consider a simple version of a two-component-spinor
Dirac-like equation (taking $\hbar=1$ throughout),
\begin{equation} \label{Dirac}
i\frac{d}{dt}|\psi\rangle=H|\psi\rangle=\frac{1}{2}(C\sigma_y+Dp_x\sigma_x)|\psi\rangle,
\end{equation}
where  $p_x$ is the momentum in the $x$
direction; $C$ and $D$ are two c-number coefficients,  $\sigma_x,\sigma_y$ are Pauli matrices in the $x$ and $y$
directions.  We first assume $C$ is time-independent, with $C=C_0>0$. Time-dependence of $C$ will be discussed much later.
We also note that the Hamiltonian $H$ here depends on $p_x$, but not on the momentum along the $y$ direction.
Such kind of Hamiltonian may be synthesized by tripod-scheme
cold atoms with a strong constraint potential in the transverse direction~\cite{MerklEPL2008},
tripod-scheme cold atoms in a two-dimensional laser setup
with mirror oscillations~\cite{ZhangPRAzb},  and trapped-ion systems
~\cite{LamataPRL,GerritamaNature}. Because the
$\sigma_x p_x$ term is the only spin-orbit coupling term, the ZB oscillation associated with
a wavepacket does not suffer from a fast damping in its oscillation amplitude~\cite{MerklEPL2008,ZhangPRAzb}.

We now examine the motion of an initial state ($t=0$) as a product state of a one-dimensional wavepacket in $x$ and an internal two-component spinor state,
\begin{equation}
\langle x|\psi(0)\rangle=G(x)\left(\begin{array}{c}1\\0\end{array}\right)e^{ip_0x},
\end{equation}
where $G(x)$ is a broad Gaussian in real space centered at $x_0=0$; $p_0$ (assumed to be zero below) is the central
momentum of the wavepacket along $x$. To better understand ZB, the dynamics can be analyzed in
the momentum space. Hence we carry out the Fourier transformation of the
wave function to get a narrow wavepacket in momentum representation (the required narrowness will be explained later),
\begin{equation}
\langle p_x|\psi\rangle=g(p_x)\left(\begin{array}{c}1\\0\end{array}\right)e^{-ix_0p_x},
\label{ip}
\end{equation}
where $g(p_x)$ is also a Gaussian as the Fourier
transformation of $G(x)$.

With the initial state specified in Eq.~(\ref{ip}),
the internal state will evolve because
it effectively experiences two ``magnetic fields": one along $y$ of strength $C_0$ and the other
along $x$ with strength $Dp_x$ for the component $p_x$ [see Eq.~(1)].
The total effective magnetic field strength is $\sqrt{C_0^2+p_x^2D^2}$
and the direction of the total magnetic field is characterized by an angle $\arctan(Dp_x/C_0)$.  To obtain
analytical results, we keep effects up to the first order of $Dp_x/C_0$, thus making approximations
$\arctan(p_x D/C_0)\approx p_x D/C_0$ and $\sqrt{C_0^2+p_x^2D^2}\approx C_0$. Physically,
our approximation is to assume that for different $p_x$ components, their effective Zeeman splitting is assumed to be almost the same, but with the internal state precessing around slightly different directions characterized by $\arctan(Dp_x/C_0)$.
In particular, the analytical solution to the Dirac-like
equation (\ref{Dirac}) can be directly obtained (for $C=C_0$), i.e.,
\begin{eqnarray} \label{evolution} \nonumber
|\psi(t)\rangle&=&\cos\left(\frac{1}{2}C_0t\right)g(p_x)\left(\begin{array}{c}1\\0\end{array}\right)e^{-ix_0p_x}\\
&+&\sin\left(\frac{1}{2}C_0t\right)g(p_x)\left(\begin{array}{c}0\\1\end{array}\right)e^{-i(x_0+\frac{D}{C_0})p_x}.
\end{eqnarray}
Here an extra $D$-dependent phase factor $e^{[-i(D/C_0)p_x]} $ in the second term arises from
the $p_x$-dependence of the spinor precession axis. The meaning of our early assumption of a narrow wavepacket in the momentum space is also clear:
the main part of the wavepacket should ensure that $|p_xD|\ll C_0$.
Considering effects due to
$(p_x D/C_0)^2$ or high orders will capture some insignificant deformation
of the wavepacket and the damping of ZB, but as demonstrated by
 our full numerical results below, these high-order effects will not
 change the essence of the physics under discussion and can be neglected here.

Interestingly, the analytical solution in Eq.~(\ref{evolution}) can be interpreted as
a superposition of two
sub-wavepackets centered at $x_0=0$ and
at $x=x_0+D/C_0=D/C_0$. The occupation probabilities of the two sub-wavepackets are given by
$\cos^2(C_{0}t/2)$ and $\sin^2(C_{0}t/2)$.  As time evolves, these two occupation probabilities
oscillate, thus giving rise to
a time-dependence of the average position denoted $\langle x\rangle$. The jittering motion is between
$x=0$ and $x=D/C_0$, with the angular frequency $C_0$. Specifically, $\langle x\rangle $ is given by
\begin{eqnarray} \nonumber
\langle x\rangle&=&\cos^2(C_0t/2)x_0+\sin^2(C_0t/2)(x_0+D/C_0)\\
&=&\frac{D}{2C_0}[1-\cos(C_0t)],
\end{eqnarray}
which is nothing but a ZB phenomenon, with the ZB frequency $C_0$, the ZB period $T_{ZB}=2\pi/C_0$,
 and the ``equilibrium point" of the oscillation at $x=D/(2C_0)$.
\begin{figure}[t]
\begin{center}
\vspace*{-0.5cm}
\par
\resizebox *{6.5cm}{4.4cm}{\includegraphics*{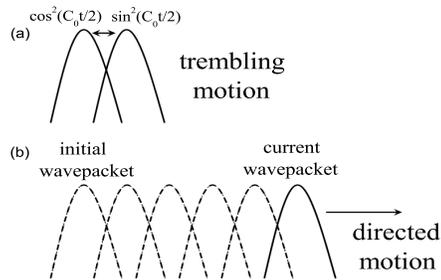}}
\end{center}
\par
\vspace*{-0.5cm} \caption{Schematic plot of wavepacket motion satisfying a Dirac-like equation.
In case (a),  a Dirac-like equation is not modulated. Cyclic population transfer between
two sub-wavepackets gives rise to ZB oscillation. In case (b), a Dirac-like equation
is periodically modulated, with the modulation frequency on-resonance with the ZB frequency.
Due to the on-resonance modulation, sub-wavepackets
successively reemerge in different positions,  yielding
directed motion.}\label{fig1}
\end{figure}
In particular, when $t=\pm\pi/C_0$, the sub-wavepacket located at $x=0$ disappears and the other
one located at $x=D/C_0$ reaches its maximal population. Figure 1(a) schematically depicts this ZB mechanism.

Given this interpretation of ZB in terms of the population cycling between two sub-wavepackets,
we next investigate what if the sign of the $\sigma_y$ term is changed at $t=\pi/C_0=T_{\text{ZB}}/2$,
with the modified Hamiltonian
given by
\begin{equation}
H=\frac{1}{2}(-C_0\sigma_y+Dp_x\sigma_x).
 \end{equation}
 At that instant,
the state of the system is given by
\begin{eqnarray}
\langle p_x|\psi(t=\pi/C_0)\rangle=g(p_x)\left(\begin{array}{c}0\\1 \end{array}\right)e^{-i(\frac{D}{C_0})p_x}.
\end{eqnarray}
Adopting the same approximations as introduced above,
we analytically obtain
\begin{eqnarray} \label{evolution2} \nonumber
|\psi(t)\rangle&=&\cos[C_0(t-\pi/C_0)/2]g(p_x)\left(\begin{array}{c}0\\1\end{array}\right)e^{-i(\frac{D}{C_0})p_x}\\
&+&\sin[C_0(t-\pi/C_0)/2]g(p_x)\left(\begin{array}{c}1\\0\end{array}\right)e^{-i(\frac{2D}{C_0})p_x}.
\end{eqnarray}
Remarkably, Eq.~(\ref{evolution2}) represents a superposition of a
sub-wavepacket located at $x=D/C_0$ and a new one at $x=2D/C_0$. The new equilibrium position of ZB is hence shifted to $x=3D/(2C_0)$.  Loosely speaking, this indicates that if
a sign change of the $C\sigma_y$ term occurs at the right moment, then the $D$-dependent extra phase factor [see Eqs. (4) and (8)] in the time-evolving state makes an opposite contribution
and consequently a sub-wavepacket at a new location emerges.
From this new solution, it is also observed that at $t=2\pi/C_0=T_{\text{ZB}}$, the occupation probability of the sub-wavepacket centered at $x=D/C_0$ will totally vanish and that of the other centered at $x=2D/C_0$ will become unity.
The average position of the system moves from $x=D/C_0$ at $t=T_{\text{ZB}}/2$ to $x=2D/C_0$ at $t=T_{\text{ZB}}$.

Repeating this strategy, i.e., abruptly changing the sign of $C$ after every interval of $T_{\text{ZB}}/2$, a directed
transport emerges from a modulated ZB. That is, if we let
\begin{equation} \label{modulation1}
C=\left\{\begin{array}{c}C_0,\
 \eta\in[0,\frac{1}{2}) \\  \\
-C_0,\ \eta\in[\frac{1}{2},1)\end{array}\right.,
 \end{equation}
 where $\eta\equiv \text{mod}(\frac{t}{T_{ZB}},1)$,
then ZB physics no longer induces oscillations. Instead, ZB is forced to become directed motion,
as schematically illustrated in
Fig. 1(b).  Certainly, modulating the $C\sigma_y$ term in a slightly different manner
may also result in directed transport in the $-x$
direction. We will not repeat the similar theoretical analysis.  In addition,
as shown below, tailored modulation of $C$ can lead to more complicated control over the wavepacket dynamics.

\section{numerical simulation}

In this section, we verify our analytical results by
numerical simulations without any approximation. Two representative modulation schemes are considered. In scheme (i),
we introduce the modulation of $C$ as described by Eq.~(\ref{modulation1}) first and then stop the modulation for  $T_\text{stop}=NT_{ZB}$ ($N$ integer), followed by the same modulation again [see Fig. 2(a)].
In scheme (ii), everything is the same except that $2T_\text{stop}/T_{\text{ZB}}$ is an odd integer [see Fig. 2(b)].
%This is
%equivalent to a restart of the modulation after $T_\text{stop}=2N\pi/|C|$.

For scheme (i), Fig. 2(c) shows that the wavepacket first moves in the $x$
direction, then undergoes ZB once the modulation is stopped, and finally the
directed motion continues after the modulation is switched on again.
The wavepacket profile in real space at different times is plotted in Fig. 2(e).
For scheme (ii), Fig. 2(d) shows that the wavepacket first moves in the same way as in the first scheme.
However, once the modulation is re-launched, the wavepacket transport
is reversed. So this behavior may be regarded as a type of ``super-ZB" oscillations (analogous to super-Bloch oscillations~\cite{Bloch1}).
As also shown in Figs.~2(e) and 2(f), in both schemes the distortion of the wavepacket profile for the shown time scale
is invisible to our naked eyes. The numerical
results here confirm our theoretical analysis, demonstrating
that ZB can be indeed converted to directed wavepacket motion in a well-controlled fashion.
\begin{figure}[t]
\begin{center}
\vspace*{-0.5cm}
\par
\resizebox *{9.5cm}{9cm}{\includegraphics*{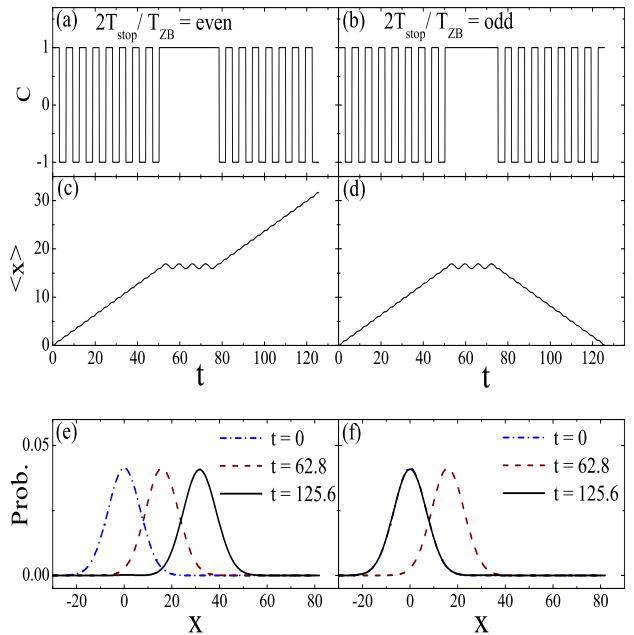}}
\end{center}
\par
\vspace*{-0.5cm} \caption{(color online) Numerical results for a modulated Dirac-like equation.
Panels (a) and (b) illustrate two schemes of modulation of the sign of system parameter $C$.  When modulation is on,
the sign of $C$ is reversed after each time interval of $T_\text{ZB}/2$ [see Eq.~(\ref{modulation1})].
The modulation off-period $T_\text{stop}$ is an integer multiple of $T_\text{ZB}$ in (a)
and a half-integer multiple of $T_\text{ZB}$ in (b).   Panels (c) and (d)
depict the time-dependence of the expectation value of $\langle x\rangle$, for modulation schemes in (a) and (b), respectively.
It is seen that in (c), the wavepacket continues its directed motion after the modulation is switched on again; but in
 (d), the wavepacket eventually reverses its directed motion.
 Panels (e) and (f) show the wavepacket profile, i.e., probability density distribution in $x$, at three different times,
 for modulation schemes in (a) and (b), respectively. It is seen that wavepacket distortion is hardly visible.
 Note that in (f), the wavepacket at $t=125.6$ is on top of that at $t=0$.  All plotted quantities and system parameters
 are in dimensionless units, with
 $C_0=D=1$.} \label{fig2}
\end{figure}

\begin{figure}[t]
\begin{center}
\vspace*{-0.5cm}
\par
\resizebox *{9.5cm}{9cm}{\includegraphics*{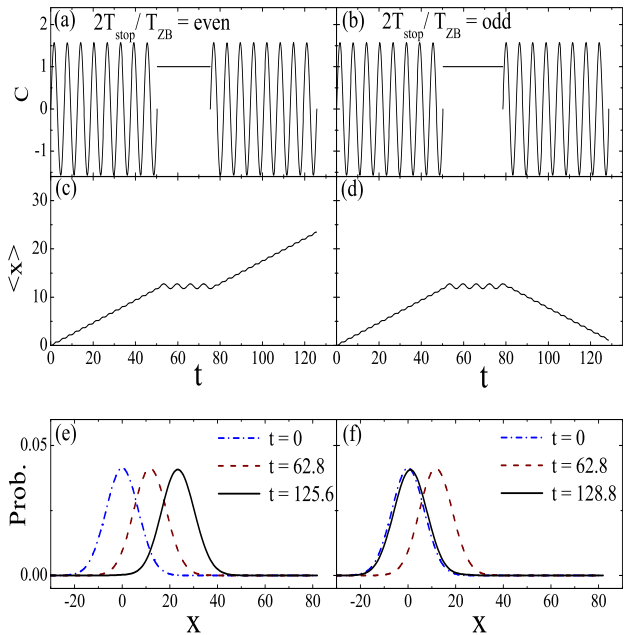}}
\end{center}
\caption{(color online) Same as in Fig.~2, except for a sinusoidal modulation of $C$, i.e., $C=C_0\sin(\omega t)$, with $\omega=1$, $C_0=\pi/2$, and $D=1$.
The pattern shown in panels (c)-(f) here are much similar to those seen in Fig.~2(c)-2(f), but the distance traveled by the wavepackets is different.}
\end{figure}

In real experiments, it should be more straightforward to introduce a smooth modulation to system parameters.
Consider then a sinusoidal modulation of $C$, $C=C_0\sin(\omega t)$.
This case is somewhat complicated because the magnitude of $C$, and hence the ZB frequency, is also time-dependent. To resolve this technical issue, we define a time-averaged effective ZB angular frequency over one half period of modulation, i.e.,
 \begin{eqnarray}
 \omega_\text{ZB}^\text{eff}\equiv \frac{\omega}{\pi} \int_{t=0}^{\pi/\omega} C_0\sin(\omega t) dt = \frac{2C_0}{\pi}.
 \end{eqnarray}
In the light of our previous interpretation of ZB oscillation, it is expected that
 after the time interval $\pi/\omega_\text{ZB}^{\text{eff}}$, one of the two sub-wavepackets has vanished and the other has an occupation probability of unity.
  In accord with our theoretical insights above, the sign of $C$ must be changing at this moment (from $0^{-}$ to $0^{+}$ or vice versa) in order to convert ZB oscillation to directed motion. This suggests the following modified resonance condition: $\omega=\omega_\text{ZB}^\text{eff} = \frac{2C_0}{\pi}$.
It should be also noted that for a sinusoidal modulation, the magnitude of $C$ can be very small during certain time windows and hence
our early treatment in terms of the first-order expansion of $D/C$ can be somewhat problematic.
Nevertheless, as seen in Fig.~3, the numerical results confirm that $\omega=\omega_\text{ZB}^{\text{eff}}$ is the required resonance condition under a sinusoidal modulation so that ZB oscillation is again converted to directed motion.  This also implies that the controlled dynamics is mainly contributed by the time segments during which
$C$ is appreciably nonzero. As seen in Fig.~3, for either even or odd $2T_\text{stop}/T_{ZB}$ ($T_{\text{ZB}}$ is now determined by $\omega_{\text{ZB}}^{\text{eff}}$), the pattern of the results presented in Fig.~3(c)-(f) is essentially the same as those seen in Fig.~2(c)-(f), with again little wavepacket distortion.  Note also that here the wavepacket transport distance per modulation cycle is less than that achieved by a mere discontinuous sign modulation of $C$.

\section{Discussion}

The Dirac-like equation in Eq. (\ref{Dirac}) can be
realized in cold-atom systems or in trapped-ion systems.  According to
the concrete experimental realization in Ref.~\cite{GerritamaNature}, the $C\sigma_y$ term
can be realized through two on-resonance laser fields, with the system parameter $C$ connected with
the phase and the strength of the laser fields. Modulation of $C$ can hence be realized by modulating the laser phase, via
a laser phase modulator or even an oscillating mirror~\cite{ZhangPRAzb}.
The ZB frequency in Ref.~\cite{GerritamaNature} can be several tens of kHz,
and the required phase modulation frequency should match this ZB frequency.  Again using the
experimental parameters in Ref.~\cite{GerritamaNature}, it can be estimated that
for a ZB frequency of $40$ kHz, a wavepacket can be transported by about $10^{-8}$m per modulation cycle.

In summary, using a simple Dirac-like equation that is already experimentally realized,
we show that ZB can be converted to directed motion via a periodic
driving of the Dirac-like equation, if the driving is on resonance with the ZB oscillation.  We hope that our results
can motivate further studies of quantum control in driven Dirac-like equations.

{\bf Acknowledgments}.
This work is supported by National Research Foundation and Ministry
of Education, Singapore (Grant No. WBS: R-710-000-008-271).
%  and by the``YIA" fund (WBS grant No.: R-144-000-195-101) (JG)
%of the National University of Singapore.

\end{document}